\documentclass[         %
aps,                    %  American Physical Society
prd,                    %  Physical Review D
nofootinbib,            %  Does not treat footnotes as references
preprintnumbers,        %
floatfix]               %  Fixes float errors
{revtex4}               %  REVTEX 4 Package used
\usepackage{graphicx}
\usepackage{graphicx,longtable}
\usepackage[dvips]{epsfig}
\usepackage{dcolumn}%Align table columns on decimal point
\usepackage{bm}% bold math
\begin{document}

\title{Unparticle Searches Through Low Energy Parity Violating Asymmetry}
\author{K. O. Ozansoy}\email{oozansoy@physics.wisc.edu}
\affiliation{Department of Physics, Ankara University, 06100 Tandogan,
Ankara, Turkey}

\date{\today}

\begin{abstract}
In this paper, we study the effects of the unparticles on the parity-violating
asymmetry for the low energy electron-electron scattering,
 $e^-e^-\to e^-e^-$. Using the data from the E158 experiment at SLAC we
extract the limits on the unparticle coupling $\lambda_{AV}$, and 
on the the energy scale $\Lambda$ at $95\%$ C.L. 
for various values of the scaling dimension $d$.  
\end{abstract}
%\medskip

%\pacs{12.60.-i, 14.80.-j}
%\keywords{unparticle sector, parity violation}

\maketitle

\section{Introduction}

The Standard Model(SM) of electroweak interactions 
has been throughly a paradigm of the elementary particle 
physics for more than three decades. It has lived through 
with the many sucessful predictions and explanations 
on the experimental observations. Many of its predictions 
has been tested sucessfully at the high energy collider 
experiments as well as the low energy experiments.
It seems next decade will be the LHC era for testing the SM predictions
and searching new physics signatures from beyond the SM. 
High energy experiments are crucial for direct observation
of the signatures from the new physics effects, however, low energy
particle physics experiments are also very important to understand
the effects in detail.  

Low energy parity-violation(PV) measurments
are very sensitive to new physics effects
around TeV energies, and are complementary to high energy
colliders. A comprehensive study on the new physics searches
through the low energy parity-violation experiments
has been given in Ref.~\cite{RamseyMusolf:1999qk}.
Atomic PV experiments, and electron-hadron, and electron-electron
scattering  PV experiments have been performed with very
high sensitivity. Amidst those experiments, PV M{\"o}ller scattering,
which is the cleanest since it is a pure leptonic process, 
is very promising to search for the new physics. There are many
works on the low energy parity violating  M{\"o}ller scattering
in the literature. For example, Refs.~
\cite{RamseyMusolf:1999qk, Derman:1979zc, Czarnecki:1995fw,Anthony:2003ub,
Anthony:2005pm, Czarnecki:2000ic, Kurylov:2003zh, Erler:2004in,
Kumar:2005cf, Kumar:1996ts, Kumar:2005rz, Musolf:1994, zykunov2005}

One of the most promising high precision PV measurmment in the
M{\"o}ller scattering was designed by the E158 experiment
at the Sanford Linear Accelerator Laboratory(SLAC)
(see, for example Refs.~ \cite{Anthony:2005pm,Anthony:2003ub,Kumar:2005cf,Kumar:1996ts,Kumar:2005rz}).
In the experiment, the PV asymmetry measurment is
obtained in the scattering of longitudinally polarized
electron beams on an unpolarized fixed target. The parity-violating, 
left-right, scattering asymmetry is defined by

\begin{eqnarray}
 A_{PV}=\frac{\sigma_R-\sigma_L}{\sigma_R+\sigma_L}
\end{eqnarray}

where $\sigma_{R(L)}$ is the scattering cross section for
incedent right(left)-handed electrons.

The high precision measurement of the PV asymmetry $A_{PV}$ has
been considered to search for the new physics effects. New physics effects 
due to $Z^\prime$ bosons, compositeness, anomalous anapole moment
effects, double charged scalar bosons, extra dimensions, etc.
have been considered regarding this parity violating asymmetry measurements,
Ref.s~\cite{Derman:1979zc, Czarnecki:1995fw, RamseyMusolf:1999qk}.

Recently, Georgi has proposed a mind-bending offspring
of a possible scale invariant sector living at very high energy
scale, Ref~\cite{Georgi:2007ek}. According to this new physics
proposal, if there is a conformal symmetry in nature it must be broken 
at a very high energy scale which is above the current energy scale of 
the colliders. Considering the idea of the Ref.~\cite{Banks:1981nn}, 
in the Ref.~\cite{Georgi:2007ek}, the scale invariant
sector is presented by a set of the Banks-Zaks operators 
${\cal O}_{BZ}$, and defined at the very high energy 
scale. Interactions of the BZ operators ${\cal O}_{BZ}$ with
the SM operators ${\cal O}_{SM}$ are expressed by the exchange of
particles with a very high energy mass scale
${\cal M}_{\cal U}^k$ in the following form

\begin{eqnarray}
\label{1}
 \frac{1}{{\cal M}_{\cal U}^k}{O}_{BZ}{O}_{SM}
\end{eqnarray}

where BZ, and SM operators are defined as
${O}_{BZ}\in {\cal O}_{BZ}$ with mass dimension $d_{BZ}$,
and ${O}_{SM} \in {\cal O}_{SM} $ with mass dimension $d_{SM}$.
Low energy effects of the scale invariant ${\cal O}_{BZ}$ fields
imply a  dimensional transmutation. Thus, after
the dimensional transmutation Eq.(\ref{1}) is given as

\begin{eqnarray}
\label{2}
 \frac{C_{\cal U} \Lambda_{\cal U}^{d_{BZ}-d}}
{{\cal M}_{\cal U}^k}{O}_{\cal U}{O}_{SM}
\end{eqnarray}

where $d$ is the scaling mass dimension of the unparticle operator
$O_{\cal U}$ (in the Refs.\cite{Georgi:2007ek, Georgi:2007si}, 
$d=d_{\cal U}$ ), and the constant $C_{\cal U}$ is a coefficient function.

Using the calculation techniques of the standard effective
field theory one can predict possible implications of the unparticles
on the particle physics phenomenology. Interactions between the unparticles 
and the SM fields have been listed by Ref.~\cite{Cheung:2007ue}. So far, many
implications of the unparticles have been studied by several works,
for example, Refs.~\cite{Georgi:2007ek,Georgi:2007si,Cheung:2007ue,Bander:2007nd,
Anchordoqui:2007dp,Hannestad:2007ys,Liao:2007bx,Davoudiasl:2007jr,
Bhattacharyya:2007pi,Ding:2007bm,Balantekin:2007eg,Chen:2007zy,Rizzo:2007xr,
Aliev:2007qw,Cakir:2007xb,Cakir:2007dz,Fox:2007sy,Iltan:2007ve,Lenz:2007nj,
Alan:2007ss,Sahin:2007pj,Grinstein:2008qk}.

In this work, we consider the fixed target $e^-e^-$ collider experiment
E158 at SLAC, to search for the unparticle physics effects.

\section{Parity Violating Asymmetry for M{\"o}ller Scattering}

\begin{figure}
\includegraphics{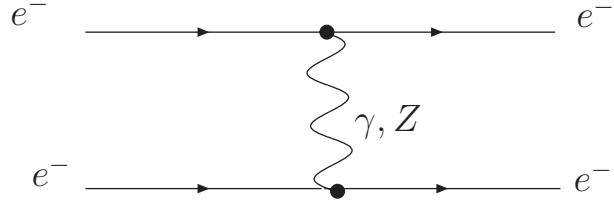}\\
\caption{ The SM contribution to the scattering amplitude
for $e^-e^-\to e^- e^-$ scattering.
\label{fig:1}}
\end{figure}

The tree level prediction of the SM for the parity-violating
asymmetry $A_{PV}$ for low energy M{\"o}ller scattering, 
Figure~\ref{fig:1}, is due to the interference between electromagnetic 
and weak neutral current amplitudes and is given by, Ref.~\cite{Derman:1979zc},

\begin{eqnarray}
A_{PV}=-\frac{G_F Q^2}{\sqrt{2}\pi\alpha(Q)}
\frac{(1-y)}{\big[1+y^4+(1-y)^{4}\big]}[1-4\sin^2\theta_W]
\end{eqnarray}

where $G_F$, and $\alpha(Q)$ are the Fermi, and the
fine structure constants, Ref.~\cite{Yao:2006px}, respectively,
and the momentum transfer is $Q^2=-q^2=-sy$ for 
$y=\frac{1}{2}(1-cos\theta)$, and the mandelstam parameter $s$.

For the fixed target experiments, the parity-violating asymmetry is
very small due to the smallness of the factor $G_F Q^2$. For 
the SLAC E158 experiment at a beam energy $\approx50$GeV and
a center of mass scattering angle $90^o$, the SM tree level prediction 
for this asymmetry is $A^{tree}_{PV}\approx 320\times10^{-9}$. However,
the electroweak radiative corrections, 
Refs.~ \cite{Czarnecki:1995fw,Erler:2004in,Kurylov:2003zh},
and the experimental precision imply about $50\%$ reduction
for the measured asymmetry. The largest radiative corrections
to $A_{PV}$ at low energies come from the $WW$ box diagrams,
and the photonic vertex, and the box diagrams, and
the $\gamma Z$ mixing and the anapole moment, Ref.~\cite{Czarnecki:2000ic}.

Following the conventions of the Ref. \cite{Czarnecki:1995fw},
one can rewrite the parity-violating asymmetry with one-loop radiative
corrections as

\begin{eqnarray}
 A_{PV}=\frac{\rho G_F Q^2}{\sqrt{2}\pi\alpha}
\frac{1-y}{1+y^4+(1-y)^4}{\cal F}_{QED}Q^{SM(eff)}_W
\end{eqnarray}

where $\rho$ is the low-energy ratio of the weak neutral and charge
current couplings, ${\cal F}_{QED}=1.01\pm0.01$ is a
QED radiation factor that includes kinematically
weighted hard initial and final state
radiation effects and y-dependent contributions from the
$\gamma\gamma$ and $\gamma Z$ box and vertex diagrams,
\cite{zykunov2005}, and the SM effective weak charge is defined as

\begin{eqnarray}
Q^{SM(eff)}_W=\Big\{ 1-4\kappa(0) s_W^2
+\frac{\alpha(M_Z)}{4\pi s_W^2}-
\frac{3\alpha(M_Z)}{32\pi s_W^2 c_W^2}(1-4s_W^2)[1+(1-4s_W^2)^2]
+F_1(y,Q^2)+F_2(y,Q^2)\Big\}
\end{eqnarray}

where $s_W=\sin\theta_W(M_Z)_{\overline{MS}},\quad
c_W=\cos\theta_W(M_Z)_{\overline{MS}}$

%\subsection{Unparticle effects}
We consider the following interactions between the
Standard Model leptons and the vector unparticles
\cite{Georgi:2007si,Cheung:2007ue}

\begin{eqnarray}
\frac{1}{\Lambda_{U}^{d-1}}
\bar e \gamma_\mu
(\lambda_{V}-\lambda_{A}\gamma_5) e O_{\cal U}^\mu+h.c.
\end{eqnarray}

where $\Lambda_U$, $\lambda_V$, and $\lambda_A$ are the
unparticle energy scale, the vector, and the axial vector
unparticle couplings, respectively.

Propagator for the vector unparticles has been given by 

\begin{eqnarray}
 [\Delta_F^{V}(q^2)]_{\mu\nu}=
\frac{A_d}{2\sin (d\pi)}[-q^2]^{d-2}
[-g_{\mu\nu}+a\frac{q_{\mu}q_{\nu}}{q^2}]
\end{eqnarray}

where $a=1$ corresponds to the Georgi original proposal, 
Ref.~\cite{Georgi:2007ek}, \cite{Cheung:2007ue},
and $a=2\frac{d-2}{d-1}$ for the conformal symmetry argument, 
Ref.~\cite{Grinstein:2008qk}, (in our calculations we assume $a=1$)
$q$ is the momentum of the unparticle and

\begin{equation}
A_d=\frac{16\pi^{5/2}}{{(2\pi)}^{2d}}
\frac{\Gamma(d+1/2)}{\Gamma(d-1)\Gamma(2d)}.
\end{equation}

\begin{figure}
\includegraphics{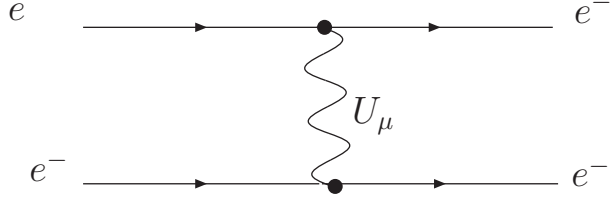}\\
\caption{ The contribution to the scattering amplitude
for $e^-e^-\to e^- e^-$ scattering from the exchange of
the vector unparticle.
\label{fig:2}}
\end{figure}

Therefore, the contribution to the transition amplitude
of $e^-e^-\to e^- e^-$ scattering from the exchange of
the vector unparticle, Figure~\ref{fig:2}, takes the form

\begin{eqnarray}
{\cal M}_{{\cal U}}= \frac{1}{\Lambda_{U}^{2d-2}}
\Big [ \frac{A_d}{2\sin(d\pi)}\Big]
\Big \{&&[-t]^{d-2}
[\bar e(p_3) \gamma_\mu
(\lambda_{V}-\lambda_{A}\gamma_5)e(p_1)]
[\bar e(p_4) \gamma^\mu
(\lambda_{V}-\lambda_{A}\gamma_5) e(p_2)]
\nonumber\\
&&-[-u]^{d-2}
[\bar e(p_4) \gamma_\mu(\lambda_{V}-\lambda_{A}\gamma_5) e(p_1)]
[\bar e(p_3) \gamma_\mu(\lambda_{V}-\lambda_{A}\gamma_5) e(p_2)]
\Big\}
\end{eqnarray}

where $t,u$ are the mandelstam parameters.

Considering the effects of unparticles the parity-violating
asymmetry can be written in the following form

\begin{eqnarray}
 A_{PV}=-\frac{G_F}{\sqrt{2}\pi\alpha}
\frac{s(1-y)}{1+y^4+(1-y)^4}
\Big[ Q^{SM(eff)}_W-\Delta Q^{U}_W\Big]
\end{eqnarray}

where the unparticle contribution is

\begin{eqnarray}
\Delta Q^{U}_{W}= \Big \{\frac{1}{2\sqrt{2}G_F}
\Big( \frac{A_d}{2\sin(d\pi)}\Big)
\frac{\lambda_{AV}}{\Lambda_U^{2d-2}}
\big [ [-t]^{d-2}+[-u]^{d-2}\big ]
\Big\}
\end{eqnarray}

where for the sake of brevity we use
$\lambda_{AV}\equiv \lambda_V \lambda_A$.

\begin{figure}
\bigskip
\includegraphics{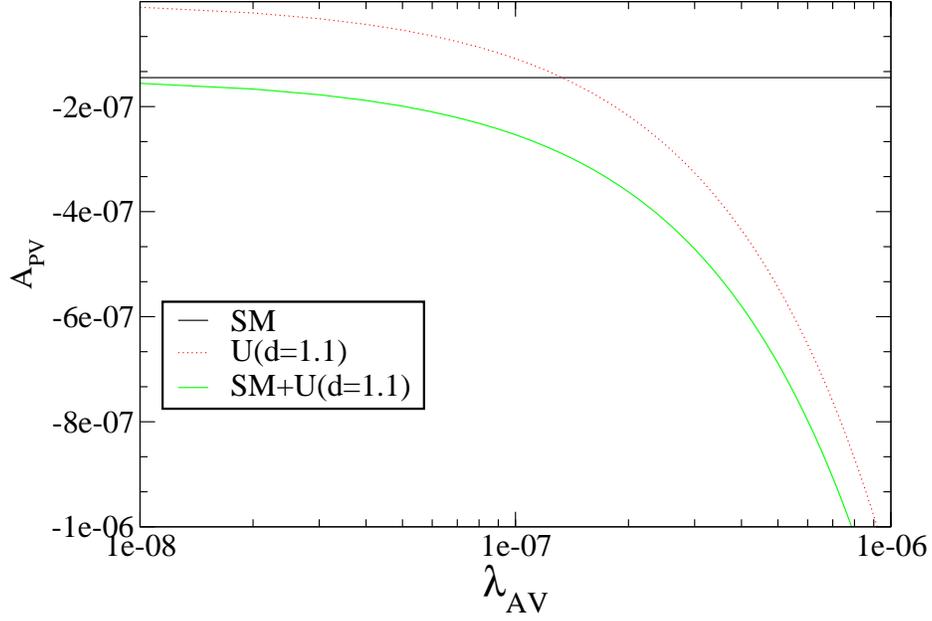}\\
\caption{ $A_{PV}$ with
respect to scaling parameter $\lambda_{AV}$.
\label{fig:3}}
\bigskip
\end{figure}

In Figure~\ref{fig:3}, 
we depict the unparticle effect on the parity-violation
asymmetry $A_{PV}$ with respect to
the unparticle coupling $\lambda_{AV}$ for  $d=1.1$ 
and $\Lambda=1000$GeV. From the figure, one can see that
for $d=1.1$ unparticle effects
are huge even the coupling $\lambda_{AV}$ is comparingly
too small. This behavior is similar for $d<1.3$.

\begin{figure}
\bigskip
\includegraphics{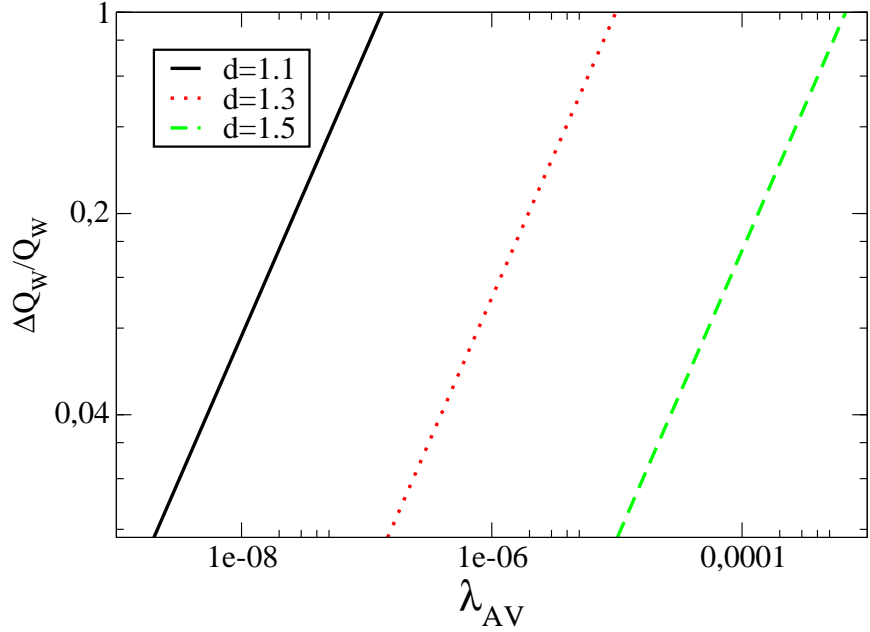}\\
\caption{ $\Delta Q^{U}_{W}$ with
respect to the unparticle coupling $\lambda_{AV}$
for $\Lambda=1000$GeV.
\label{fig:4}}
\bigskip
\end{figure}

\begin{figure}
\bigskip
\includegraphics{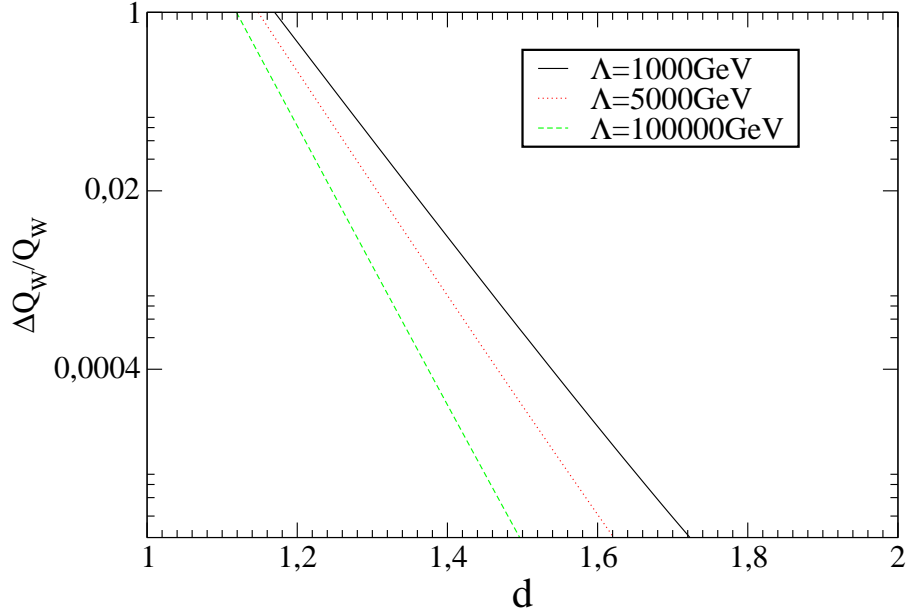}\\
\caption{ $\Delta Q^{U}_{W}$ with
respect to scaling parameter $d$.
\label{fig:5}}
\bigskip
\end{figure}

In Figure~\ref{fig:4}, and Figure~\ref{fig:5}
we depict $\Delta Q^{U}_{W}/Q_W$ with
respect to the unparticle coupling $\lambda_{AV}$, 
and the scaling parameter $d$, respectively.
In Figure~\ref{fig:4} we assume $\Lambda=1000$GeV for different
values of the scaling parameter $d$.
In Figure~\ref{fig:5} we assume $\lambda_{AV}=10^{-6}$
for two different unparticle energy scale $\Lambda_U$.
According to those figures, 
it is clearly seen that the unparticle effects 
for $d<1.3$ are very significant for the given configurations
of the $(\lambda_{AV},\Lambda)$.

\section{Results and Discussions}

According to the latest report given
by the E158 Collaboration, Ref.~\cite{Anthony:2005pm},
the combined result for the parity-violating asymmetry is

\begin{eqnarray}
 A^{EXP}_{PV}=-131\pm14(stat)\pm10(syst)ppb
\end{eqnarray}

where the average values of the kinematical variables
are $Q^2=0.026\text{GeV}^2$, and $y=Q^2/s\approx0.6$.
Using this experimental result, for fixed values of the
scaling dimension $d$, and assuming $\Lambda_U=1000GeV$,
we extract the upper limits on the unparticle coupling
$\lambda_{AV}$.
In the calculations, we use the standard
chi-square analysis for the following $\chi^2$ function

\begin{eqnarray}
 \chi^2=\big [\frac{{A_{PV}^{EXP}-A_{PV}^{SM+{\cal U}}
(\lambda_{AV})}}{\delta}\big ]^2
\end{eqnarray}

where $\delta=\sqrt{\delta_{syst}^2+\delta_{stat}^2}$
For the one sided chi-square analysis, we assume $\chi^2\geq 2.7$
which corresponds to the $\%95$ C.L. limits.
Our results are given in the Table \ref{tab1}.

\begin{table}[tbp!]
\bigskip
{\caption{Upper limits on the $\lambda_{AV}$ for various values of
the scaling parameter $d$. Here, we assume $\Lambda_U=1000$GeV.
\label{tab1}}}
\begin{ruledtabular}
\begin{tabular}{rl}
d & $\lambda_{AV}$\\
\hline
%d=1.01  & $4.8\times10^{-9}$ \\
d=1.1   & $1.3\times10^{-8}$ \\
%d=1.2   & $2.9\times10^{-7}$ \\
d=1.3   & $1.0\times10^{-6}$ \\
%d=1.4   & $2.1\times10^{-5}$ \\
d=1.5   & $6.7\times10^{-4}$ \\
%d=1.6   & $1.3\times10^{-3}$ \\
d=1.7   & $3.8\times10^{-3}$ \\
%d=1.8   & $6.2\times10^{-2}$ \\
d=1.9   & $1.2\times10^{-1}$
\end{tabular}
\end{ruledtabular}
\end{table}

\begin{figure}[tbp!]
\bigskip
\includegraphics{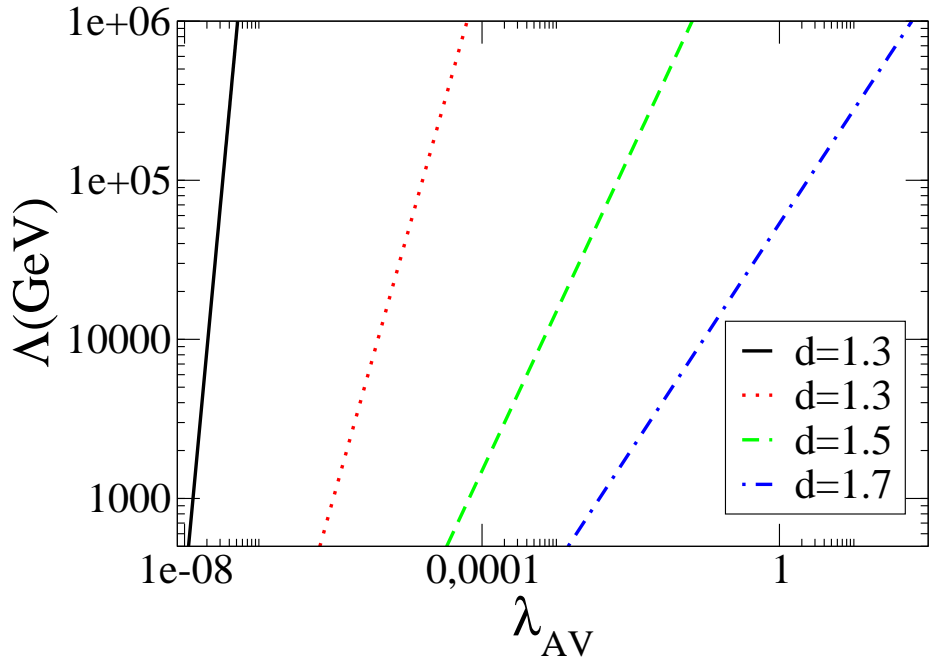}
\caption{  Upper limits on the scalar unparticle
coupling $\lambda_0$ depending on $\Lambda_{\cal U}$
\label{fig:6}}
\end{figure}

Since unparticle contribution to the parity violating
asymmetry is proportional $\lambda_{AV}/\Lambda_U^{2d-2}$,
using the above limits for $\lambda_{AV}$ one can plot
the parameter space of $\lambda_{AV}$ versus
$\Lambda_U$, Figure \ref{fig:6}. In the figure,
right hand side of each curve is ruled out according to
the $95\%$C.L. analysis for corresponding scaling dimension $d$.

For a comparision the limits on the vector or the axial-vector unparticles 
from the literature are summarized in the Table~\ref{tab2}; for the lower
limits on the unparticle energy scale $\Lambda_U$ couplings
$\lambda_{V,A}=1$ are assumed; for the upper limits on the couplings
$\Lambda_U=1TeV$ is assumed. One can easily see that our results are very 
stringent and comparable the most stringent limits excisting in the literature.

\begin{table}[htp!]
{\caption{Limits on unparticles from the literature
\label{tab2}}}
\begin{ruledtabular}
\begin{tabular}{clll}
Experiment & limits for various $d$ values  \\
\hline
LEP (\cite{Bander:2007nd})
& $d=1.1$ & $d=1.5$ & $d=1.9$ \\
$e\mu$  & $\Lambda_U(TeV)>9.1\times10^{14}$  & $\Lambda_U(TeV)>61$
& $\Lambda_U(TeV)>3.7$  \\
$e\tau$ & $\Lambda_U(TeV)> 9.7\times10^{12}$  & $\Lambda_U(TeV)>25$
& $\Lambda_U(TeV)>2.2$  \\
$eq$    & $\Lambda_U(TeV)>2.8\times10^{14}$  & $\Lambda_U(TeV)>52$
& $\Lambda_U(TeV)>3.5$  \\
$eb$    & $\Lambda_U(TeV)>1.9\times10^{11}$  & $\Lambda_U(TeV)>4.5$
& $\Lambda_U(TeV)>0.45$ \\
\hline
$e^-e^+\to\gamma X$ (\cite{Cheung:2007ue})
 & $d=1.4 $ & $d=1.6$ & $d=1.8$ \\
{} & $\Lambda_U(TeV)>660$ & $\Lambda_U(TeV)>23$ & $\Lambda_U(TeV)>4$ \\
\hline
Atomic parity violation (\cite{Cheung:2007ue}) through
$eedd$ & $d=1.4$ & $d=1.5$ & $d=1.9$ \\
{} & $\Lambda_U(TeV)>100$ & $\Lambda_U(TeV)> 30$ & $\Lambda_U(TeV)> 2$  \\
\hline
Atomic parity violation (\cite{Cheung:2007ue}) through
$eeuu$ & $d=1.4$ & $d=1.5$ & $d=1.9$ \\
{} & $\Lambda_U(TeV)> 100$ & $\Lambda_U(TeV)> 25$ & $\Lambda_U(TeV)> 1$  \\
\hline
Atomic parity violation (\cite{Bhattacharyya:2007pi}) for
 $\lambda^e_A=1$,$\lambda_V^{d,u}=1$ & $d=1.1$ & $d=1.5$ & $d=1.8$ \\
{} & $\Lambda_U(TeV)>6$ & $\Lambda_U(TeV)>2$ & $\Lambda_U(TeV)>1$ \\
\hline
$(g-2)_{e}$ (\cite{Liao:2007bx}, \cite{Rizzo:2007xr}) & {} & d=1.5  \\
{}& {} & $\Lambda_V(TeV)>37$ \\
\hline
$(g-2)_e$ (\cite{Liao:2007bx}, \cite{Rizzo:2007xr})
 & {} & d=1.5  \\
{}& {} & $\Lambda_A(TeV)>146$ \\
\hline
$(g-2)_{\mu}$ (\cite{Rizzo:2007xr}) & d=1.5 & d=1.6 \\
{} & $\Lambda_V(TeV)>1000$ & $\Lambda_V(TeV)>10$ \\
\hline
$(g-2)_{\mu}$ (\cite{Rizzo:2007xr})
 & d=1.5 & d=1.6  \\
{}& $\Lambda_A(TeV)>100$ & $\Lambda_A(TeV)>1$\\
\hline
Invisible positronium decays (\cite{Liao:2007bx})  & {} & d=1.5  \\
{}& {} & $\Lambda_V(TeV)>4.3\times10^{5}$\\
\hline
Invisible positronium decays (\cite{Liao:2007bx})
 & {} & d=1.5  \\
{}& {} &  $\Lambda_A(TeV)>5.1\times10^{2}$\\
\hline
Atomic PV through Ba, and Yb isotope chains (\cite{Ding:2007bm})
& $d=1.1$  & $d=1.5$ & $d=1.9$ \\
{} & $\lambda_{AV}<10^{-3}$& $\lambda_{AV}<2\times10^{-2}$
& $\lambda_{AV}<1.1\times10^{-1}$\\
\hline
Low energy $\nu e$ scattering (\cite{Balantekin:2007eg})
& d=1.1 & d=1.5 & d=1.9\\
{}& $\lambda_V<6.2\times10^{-6}$ & $\lambda_1<9.1\times10^{-3}$
& $\lambda_1<8.1$\\
\hline
Invisible decays of Z (\cite{Chen:2007zy}) & {} & d=1.3 & d=1.5\\
{} & {} & $\lambda_V<0.049$ & $\lambda_V<0.1$\\
{} & {} & $\Lambda_V(TeV)>10^4$ & {}
\end{tabular}
\end{ruledtabular}
\end{table}

As a conclusion, we study the unparticle effects on the parity violating
asymmetry $A_{PV}$ in the low energy electron-electron scattering.
We show that the parity-violating asymmetry $A_{PV}$ measurments,
which are complementary to the high energy collider experiments to seek
for the new physics effects, give very stringent limits on unparticles,
especially for the values $d<1.3$.

We would like to remark that the recent proposal on the possibility
to perform a new measurement at the Jefferson laboratory can 
potentialy achieve a factor of 5 improvement over the result
of the E158 measurement, Ref.\cite{Kumar:2005rz}. Therefore, such an 
improvement would give better understanding of the new physics effects,
and can be used to put more stringent limits on the new physics
scenarios, such as the unparticle physics.

\section*{ACKNOWLEDGMENTS}

It is a pleasure to thank M. Ramsey-Musolf, F. Petriello, K. Kumar,
and B. Balantekin for helpful conversations and discussions on this 
work. I also would like to thank to the members of the Nuclear Theory 
Group of University of Wisconsin for their hospitality.

\end{document}